# Strain Dependent Spin Hall Magnetoresistance in the Multiferroic Antiferromagnet BiFeO$_3$

D. Sando[1,2,3,*,+], S. Chen[4,+], O. Paull[1,3], B. Xu[5], J. J. L. van Rijn[4], C. Xu[6], S. Xu[5], F. Appert[7], J. Juraszek[7], L. Bellaiche[3,5], V. Nagarajan[1,3], T. Banerjee[3,4,*]

[1]School of Materials Science and Engineering, UNSW Sydney, Kensington 2052, Australia
[2]School of Physical and Chemical Sciences, University of Canterbury, Christchurch 8041 New Zealand
[3]Australian Research Council Centre of Excellence in Future Low Energy Electronics, Australia
[4]University of Groningen, Zernike Institute for Advanced Materials, 9747 AG Groningen, The Netherlands
[5]Jiangsu Key Laboratory of Thin Films, School of Physical Science and Technology, Soochow University, Suzhou 215006, China
[6]Department of Physics and Institute for Nanoscience and Engineering, University of Arkansas, Fayetteville, AR, USA
[7]Univ Rouen Normandie, INSA Rouen Normandie, CNRS, Normandie Univ, GPM UMR 6634, F-76000 Rouen, France

[+]These authors contributed equally

Email: daniel.sando@canterbury.ac.nz , t.banerjee@rug.nl

*Abstract*

The spin Hall magnetoresistance (SMR) of epitaxial BiFeO$_3$ thin films is investigated. SMR consistent with ferromagnetic interfacial states for BiFeO$_3$ films fabricated on (001) SrTiO$_3$ (R' BFO) and LaAlO$_3$ (T' BFO) substrates is found, albeit with different temperature dependencies. For T' BFO, the SMR is enhanced at room temperature, and decays with reduced temperatures. By contrast, R' BFO shows a monotonic decrease in SMR response with increasing temperature, mirroring the trend of a weak ferromagnet. Density functional theory shows that this difference originates from the coupling of the applied magnetic field to oxygen octahedral rotation (R') and spin (T') degrees of freedom.

Bismuth ferrite (BiFeO$_3$ - BFO), a widely studied multiferroic oxide, has served as a fertile playground for understanding electric field manipulation of magnetism [1–3]. The sensitivity of BFO's crystal structure to epitaxial constraint is well established [4,5]: large in-plane compressive strain induces a rhombohedral-like (R') to tetragonal-like (T') phase transition [6], with concomitant modification of ferroelectric, magnetic, and electromechanical properties [7]. Spin-lattice coupling effects in BFO have also been widely studied [8–11].

More recently, BFO, an antiferromagnet (AFM) with a net weak FM [12], has also garnered significant attention for its magnonic response [13,14]. The electrical generation and propagation of magnons forms a major thrust in developing antiferromagnet-based spintronics [15]. Parsonnet *et al*. recently studied magnons in epitaxial BFO thin films [16]. Using the spin Seebeck effect, they demonstrated transport of magnons whose propagation was hindered by repeated scattering at ferroelectric/magnetic domain walls, revealing that domain morphology (of both magnetic and ferroelectric order parameters) and magnetic anisotropy govern magnon propagation. As mentioned previously (and showcased in **Fig. 1**) epitaxial strain induces prominent changes to the crystal structure in BFO. Given that the ferroelectric polarization and the magnetic structure is closely linked to the crystallographic phase, one should expect T' BFO to have significantly different magnonic response compared to R' BFO. Moreover, the manifestation of the magnetic anisotropy through changes in the magnetoresistance (MR) under a range of external conditions such as temperature and magnetic field has significant device implications [17].

In this Letter, we demonstrate strain-induced phase tuning of the SMR response in epitaxial BFO layers. We show that the surface spin configurations are dramatically different for T' vs. R' ferroelectric structural phase, which can be electrically read out using spin Hall magnetoresistance (SMR). Whilst surprisingly the SMR suggests a ferromagnetic state for both R' and T' BFO, the temperature dependencies show stark differences. In T' BFO, the SMR signal is enhanced at room temperature, and it decays with reduced temperatures. By contrast, in R' BFO the SMR response decreases upon increasing temperature, essentially mirroring the temperature trend of the weak FM. We understand these differences in the framework of the weak magnetic moments in the two BFO phases, whereby in T' BFO the moment induced directly by the field gives the SMR response, while in R' BFO, the weak magnetic moment mediated by the **L** vector dominates the behavior. Our results demonstrate that (i) strain and/or phase engineering can be harnessed to modulate the SMR signal and its temperature dependence, ii) mesoscale BFO devices show promise for spintronic and magnonic technologies, and iii) it is critical to account for weak FM in antiferromagnets for the development of AFM spintronic devices.

SMR is a phenomenon that can read out the surface spin configuration of magnetic insulators (MIs) [18–21], making it a powerful interface-based technique to fingerprint magnetic anisotropy by electrical means. A charge current travelling through a heavy metal (such as Pt) generates a transverse spin current through the spin Hall effect, yielding spin accumulation at the Pt surface [**Fig. 1 (a)**]. At the interface between the MI and Pt, the spins of the Pt electrons transfer their angular momentum, dependent on the magnetization orientation of the MI layer. Consequently, the spin current is absorbed (reflected) when the interfacial magnetic moments are oriented perpendicular (parallel) to the spin accumulation. The absorbed spin currents exert a spin transfer torque on the adjacent MI layer and dissipate, resulting in a detectable increase in the Pt resistivity through the inverse spin Hall effect. An applied magnetic field modifies the orientation of the magnetization in the MI layer, thereby modulating the resistance of the Pt layer. **Figure 1(b)** depicts the typical measurement geometry. In ferromagnets (FM) with small anisotropy, the magnetization vector **M** follows the rotating external magnetic field, resulting in an angular dependence of the resistance, defined as positive SMR [**Fig. 1(c)**] [22]. In contrast, in some AFM insulators, negative SMR has recently been observed [23,24]. Here, the Néel vector **L** remains orthogonal to the applied magnetic field, which, when compared to the positive SMR case, leads to a 90-degree phase shift in the angular dependence. The phase alignment in SMR can thus be exploited as a fingerprint to distinguish FM and AFM interfaces.

Given that SMR is an interface-dominated effect, it offers several advantages over traditional volume averaged techniques such as neutron diffraction [25–28] Mössbauer spectroscopy [14,29,30], and Raman spectroscopy [13,14,31,32] when studying thin films. SMR-based multiferroic devices promise a new paradigm of electrically controlled spintronics, since 1) SMR can be applied for the understanding of transport of magnons [24], 2) SMR being an interface effect, it should be highly susceptible to strain-induced changes in spin order; 3) it is an all-electrical probe, attractive for low-energy spintronic devices.

Our first insights come from density functional theory (DFT) calculations which we used to determine the preferred structural phases, magnetic ground states and weak ferromagnetism (wFM) of BFO at zero applied magnetic (H) field, at 0 K [**Figs. 1(d-f)**]. The calculated energy of G-type and C-type AFM ordering as a function of strain [**Fig. 1(d)**] shows that under moderate strain (from -4 % to +2 %), where R' BFO is stable, the AFM ground state is G-type. Moreover, the **L** vector is parallel to [$1\bar{1}0$] (not shown) (pseudocubic indices are used throughout). On the other hand, at high compressive strain (-4 % to -6 %, where T' BFO is stable), G-type and C-type AFM are essentially degenerate; however, for consistency with experiments on T' BFO [6], in the following we impose G-type AFM order.

The calculated components of the weak magnetization as a function of strain are shown in **Fig. 1(e)**. Under -1.5 % strain, corresponding to BFO grown on SrTiO$_3$ (STO), $M_x = M_y = +0.08\ \mu_B$, while $M_z = -0.14\ \mu_B$, yielding an **M** vector pointing approximately along [$11\bar{2}$]. By contrast, under < -4.5% compressive strain (BFO on LaAlO$_3$ - LAO), $M_x = M_y = 0$, and the z component is very small (+0.04 $\mu_B$), yielding **M** // [001]. The direction of these **M** vectors is consistent with the rule that the wFM is proportional to the cross product between the AFM vector and the anti-phase oxygen octahedral tilting angle [**Fig. 1(f)**] [33,34].

Since our DFT results find the magnetism for the T' and R' phases to be markedly different, it presents a unique opportunity to demonstrate hitherto unseen strain tuning of SMR, of which **Fig. 2** is experimental demonstration. We grew ~50 nm thick (001) BFO films by pulsed laser deposition onto STO and LAO substrates using conditions reported elsewhere [35–37]. We are uniquely able to fabricate phase pure T-BFO films up to 70 nm in thickness, with no trace of mixed phase striations [32]. This capability allows us to establish the clear contrast in behaviors of the T' and R' phases of BFO – and hence the remanent strain state – for a fixed film thickness of 50 nm, without complications from mixed R'/T' BFO. In the T' BFO, 50 nm is thin enough to be fully strained and avoid complications from dislocations and other ferroelastic domain formations, but not so low as to suffer from interface-driven size scaling effects. The uniform strain state moreover ensures a spatially homogeneous magnetic order that is picked up in a device geometry. Conversion electron Mössbauer spectroscopy [12] on comparable samples with 100% $^{57}$Fe enrichment shows magnetic hyperfine sextets with an intensity ratio of peaks 2 and 3, $R_{23}$ close to 4, implying that the AFM vector is confined to the (001) plane [**Fig. 1 (g,h)**], consistent with **L** // [$1\bar{1}0$] (Refs. [7,14,38]). The sextets for the two phases do not show asymmetry characteristic of a cycloidal modulation of the spins [14]. While the spectrum of R' BFO presents a hyperfine field close to that of bulk BFO ($B_{hf}$= 48.8 T), the spectrum of T' exhibits broader lines and a smaller hyperfine splitting with $\langle B_{hf}\rangle$ = 29.5 T, due to reduced magnetic interactions near the critical transition temperature of ~350 K [7].

We next performed temperature dependent SMR experiments. Hall bars of 7 nm thick Pt were patterned by e-beam lithography onto the T' and R' BFO samples. Full measurements of the longitudinal and transverse resistance as a function of various rotation angles of the applied H field revealed a much cleaner response for the transverse signal when compared to the longitudinal one (not shown). We focus therefore on the transverse resistivity $\rho_{xy}$ and its dependence on temperature. **Figure 2(a)** shows the dependence of $\rho_{xy}$ on the angle $\alpha$ [defined in **Fig. 1(b)**] at temperatures 50-370 K for T' BFO. The trend in $\rho_{xy}$ is consistent with positive SMR, reminiscent of a FM response. Fitting the data to a model comprising two sinusoidal functions (**Figs. S1**, **S2**) we extract the magnitude of the SMR $\Delta\rho_{xy}$ as a function of temperature [**Fig. 2(b)**]. SMR is not detected at 50 K, while for temperatures 200 K up to 370 K (the maximum for our setup), it monotonically increases.

By contrast, in the R' BFO sample, the 180-deg periodic α-dependent oscillations in $\rho_{xy}$ are very clear at 5 K, showing a positive SMR (FM-like) response, but the temperature dependence is drastically different: $\Delta\rho_{xy}$ diminishes in amplitude upon increasing temperature [**Fig. 2(c)**]. The key finding of *positive* SMR, along with the opposite dependence of the SMR amplitude with temperature, is rather intriguing and indicative of the different origin of magnetic ordering in T' BFO and R' BFO films. The observed positive SMR in both cases, even though they are nominally (spin canted) AFMs, could arise from various sources. Since we observe positive SMR, it appears (at first glance) unlikely that the magnetic field is purely manipulating **L**, since in this case the sinusoidal modulation of $\rho_{xy}$ would give negative SMR. On the other hand, even if the applied H is insufficient to manipulate **L**, it can increase the canting angle of the wFM of the two AFM sublattices, yielding a positive SMR.

We intuitively rationalize the behavior of T' BFO with temperature as follows. Since the external magnetic field is applied in plane, SMR is sensitive only to the in-plane magnetization component [18]. According to **Fig. 1(e)**, at low temperature, T' BFO intrinsically has a vanishingly small value of $M_x$ and $M_y$, we thus observe no SMR signal. Upon increasing temperature to 200 K, the applied field induces a non-zero moment, giving a detectable SMR response. Upon further increase of temperature, the magnitude of the moment induced by the applied field increases, which amplifies the observed SMR response and $\Delta\rho_{xy}$. Essentially, the magnetic susceptibility of T' BFO increases with temperature, thus giving a larger induced **M** under applied field, consistent with the monotonic increase in the SMR from 200 to 370 K.

Next, we describe the temperature dependence of the SMR signal for R' BFO, which is categorized into two regimes: T > 80 K and T < 80 K. The SMR amplitude shows a monotonic decrease with increasing temperature, but, interestingly, below 80 K, it sharply increases. Ignoring for now this sharp upturn, the SMR appears to be consistent with the general downwards trend of the total wFM of BFO [39,40] upon increasing temperature.

We rationalize this behavior as follows. In BFO, **M** is related to **L** and the octahedral tilting pseudovector **ω** through the following relationship:

$$\mathbf{M} \propto \mathbf{L} \times \boldsymbol{\omega}. \tag{1}$$

Here, **ω** characterizes the octahedral tilting of the unit cell, with the direction giving the axis about which the octahedra rotate in antiphase, and its magnitude giving the rotation in radians [33]. In R-BFO under moderate strain, **ω** ~ // [111]. Now, assume that an applied **H** field causes **L** to orient orthogonal to the applied field (see Ref. [31]). For example (neglecting the magnitude of these vectors), if **H** // [100], then **L** // [010], and if **ω** // [111] [**Fig. 1(e)**], then, according to Eq. (1), **M** // [10$\bar{1}$]. Here, the in-plane component of **M** is *parallel with the applied field* **H**. If the rotating applied magnetic field causes **L** to rotate synchronously orthogonal to **H**, then the induced **M** (according to Eq. 1) will also oscillate (albeit with some slight angle-dependent amplitude due to varying proportions of in-plane and out-of-plane **M** components).

The key observation is that if **M** is indeed dictated by Eq. 1 (that is, mediated by **L**), then its in-plane component will always be aligned with the applied **H** field – giving rise to a seemingly *ferromagnetic* SMR response. Now, if the total magnetic moment of BFO shows a decrease of magnitude upon increasing temperature [39], then one should anticipate the same decreasing trend in SMR of R' BFO, as observed in **Fig. 2(d)**. Note that for T' BFO, this line of reasoning does not hold, as the octahedral rotation patterns are different [i.e., **ω** // [110] from **Fig. 1(f)**], and the total magnitude of **M** for T' BFO is much smaller [**Fig. 1(e)**]. For T' BFO upon increasing temperature, the moment *induced directly by the applied field* (rather than mediated by **L**) dominates the SMR response.

Briefly summarizing the SMR results for the two sample types, we observe SMR for T' BFO that is sizeable at 300 K but negligible at lower temperatures. In contrast, the SMR for R' BFO is strongest at 5 K and monotonically decreases upon increasing temperature. In the latter case, there appears to be two temperature regimes, with a possible transition at around 80 K. To gain further insight into the SMR response and search for possible origins for the two regimes for R' BFO, we performed field-dependent measurements.

Figure 3 presents field dependent measurements at 300 K for T' BFO, and at 5 K for R' BFO [**Fig. 3(a,c)**]. Field-dependent data were also taken for R' BFO at 150 K (not shown). For both samples, the extracted $\Delta\rho_{xy}$ shows a linear dependence with field – specifically T' BFO at 300 K [**Fig. 3(b)**] and R' BFO at 5 K and 150 K [**Fig. 3(d)**]. According to SMR theory [18], at a given measurement temperature, the SMR signal increases linearly with the induced magnetic moment (or correlated to the spin fluctuations). The observed linear field dependence of SMR for both R' and R' BFO has three implications. First, it cannot result from domain reorientation, which would give a quadratic field dependence [19], second, no spin flop transition occurs, as this would yield an abrupt change in the sign of SMR when **L** reorients with the field, and third, no transition from cycloidal-pseudocollinear AFM takes place, as this would also presumably be manifest in the slope of $\Delta\rho_{xy}$ vs. H. In other words, an applied magnetic field as small as 1 T can manipulate the **L** vector (to remain orthogonal to the field), but upon increasing field, the only change is the magnitude of the induced magnetic moment.

We conclude with a brief discussion. Although BFO is well-known to be a canted cycloidal antiferromagnet (Refs. [3,12]), our observed positive SMR in all cases suggests that this experiment probes primarily the canted moment. However, R' BFO and T' BFO show clearly different temperature dependencies, with the SMR enhanced at room temperature for T' BFO. These temperature dependencies arise due to differences in the weak moment induced by the applied field. In R' BFO, the weak FM follows the applied field, mediated through the **L** vector. In contrast, for T' BFO, the vanishingly small in plane FM is not locked to **L**, so the moment induced by the applied magnetic field dominates the SMR response, and increases monotonically with temperature. Nevertheless, the increased SMR at low temperatures for R' BFO warrants further study. The structural STO phase transition at 105 K may induce a change in the octahedral rotation pattern and/or magnetic order of BFO, a spin-glass transition with blocking temperature ~50 K occurs [41], or that below 80 K, BFO develops a cycloidal order [42] which has no intrinsic net FM and may be more sensitive to an applied field. Further SMR experiments using R' BFO grown on a substrate without a low temperature structural transition (like $NdGaO_3$, Ref. [43]), or low temperature NV-center scanning magnetometry [10] could shed important insight on this issue.

*Author Contributions*

T.B., D.S., and V.N. conceived and supervised the study. O.P. fabricated the films and carried out structural characterization. S.C. patterned Hall bars and carried out SMR measurements and analysis with input from J.J.L.v.R and T.B. C.X., B.X., and S.X. performed calculations under the supervision of L.B. F.A. and J.J. performed Mössbauer spectroscopy measurements and analysis. D.S and V.N. wrote the manuscript with input from S.C., T.B., B.X. and L.B. All authors discussed the data and contributed to analysis and feedback.

*Acknowledgements*

This research was partially supported by the Australian Research Council (ARC) Centre of Excellence in Future Low-Energy Electronics Technologies (project no. CE170100039) and funded by the Australian Government. This work is realized using the facilities available at NanoLab NL. S.C. acknowledges funding support from the European Union Horizon 2020 research and the innovation program under grant agreement no. 696656. J.J.L.vR. acknowledges financial support from a Dieptestrategie grant (2019), Zernike Institute for Advanced Materials. S.X. and B.X. acknowledge financial support from National Natural Science Foundation of China under Grant No. 12074277. LB acknowledges support from the Vannevar Bush Faculty Fellowship (VBFF) from the Department of Defense and Award No. DMR-1906383 from the National Science Foundation Q-AMASE-i Program (MonArk NSF Quantum Foundry). F.A. and J.J. acknowledge the support of the Région Normandie and European Regional Development Fund of Normandy (ERDF) through the MAGMA project. We are grateful to M.M. Seyfouri and T. Musso for assistance with figures. We thank M. Bibes for discussions.

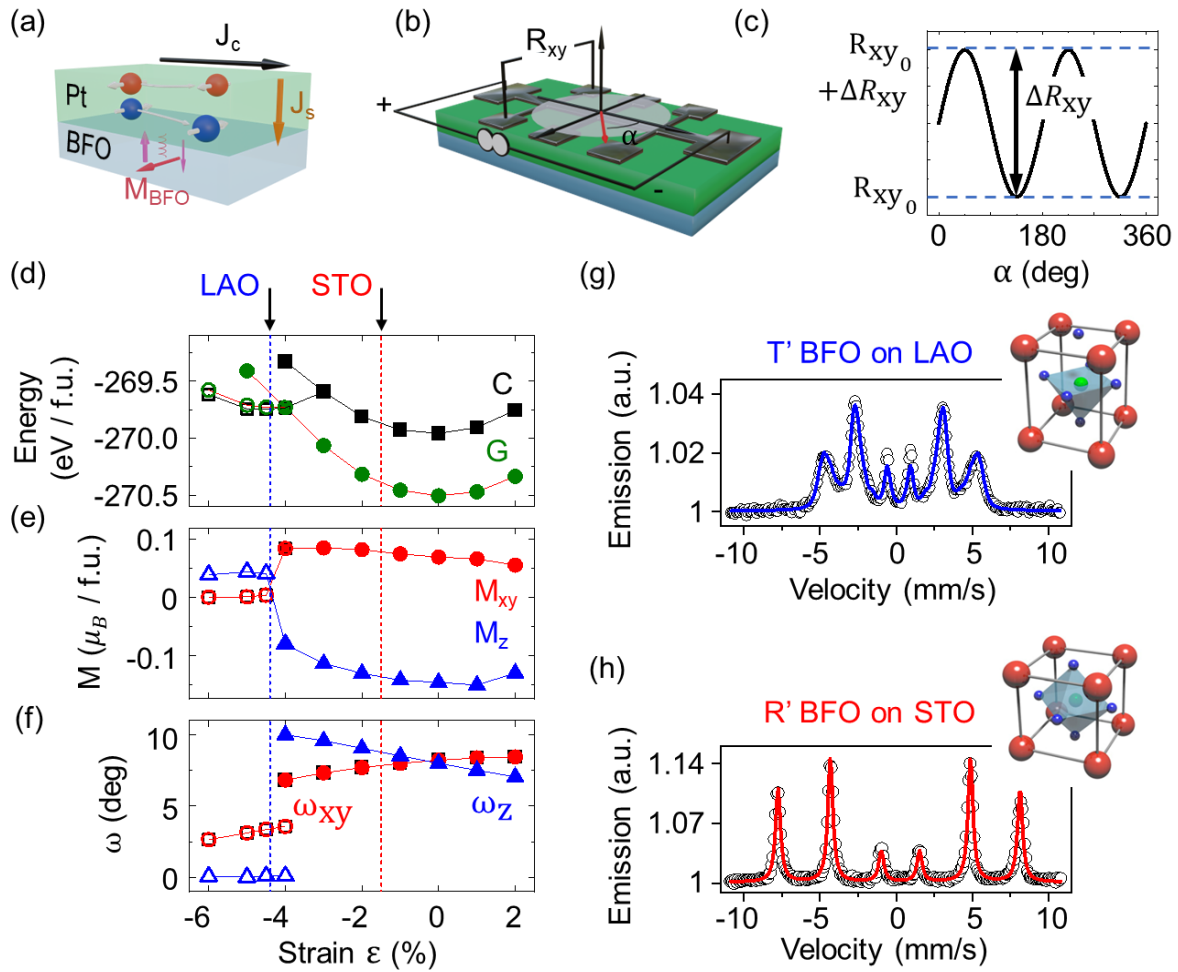

**Figure 1.** (a) Principle of SMR, (b) measurement of transverse resistivity $\rho_{xy}$ with varying applied magnetic field, (c) dependence of the transverse resistivity on angle $\alpha$; note that $R_{xy_0}$ is defined as the minimum value of resistivity. Density function theory calculated energy (d), magnetization components (e), and octahedral rotations (f) of various magnetic structures in T' and R' BFO phases, as a function of strain. (g,h) conversion electron Mössbauer spectra (at 300 K) of T' and R' BFO phases.

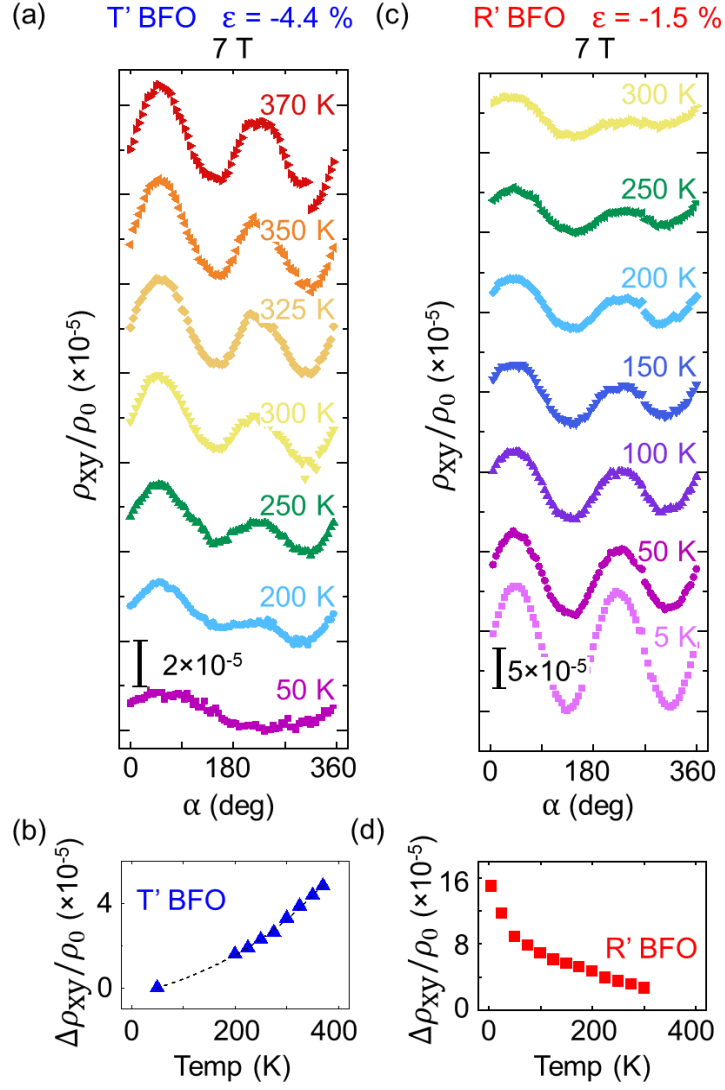

**Figure 2.** (a) T' BFO SMR response and (b) temperature dependence of the normalized Δ$\rho_{xy}$. (c) R' BFO SMR response and (d) extracted Δ$\rho_{xy}$. The additional 360° periodicity with the 180° SMR periodicity comes from a small out-of-plane Hall component related to slight misalignment of the in-plane rotator (Ref. [44]).

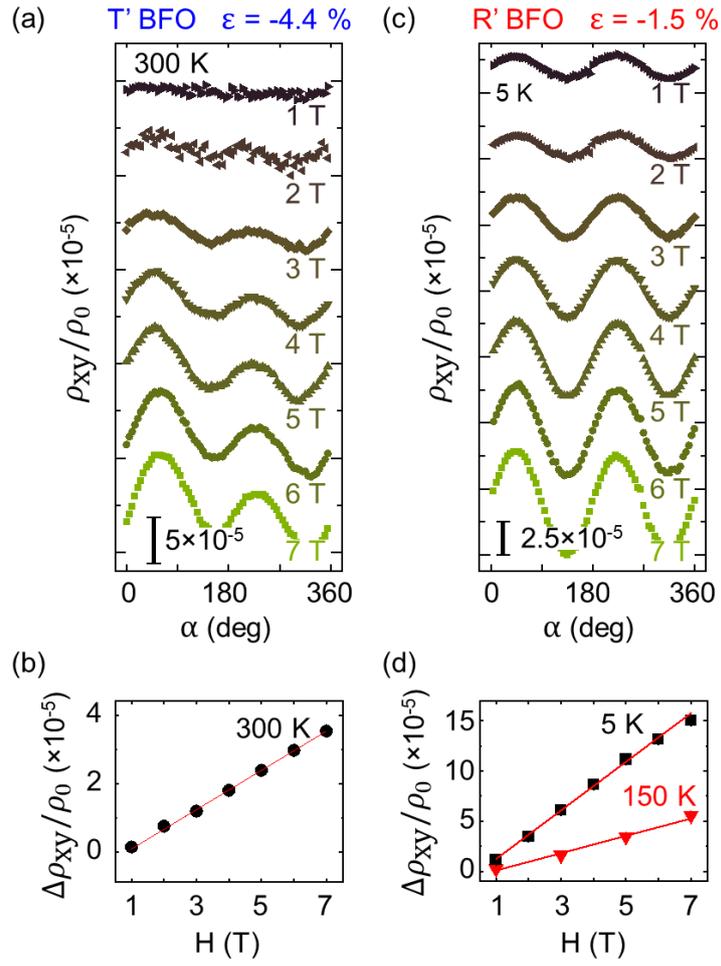

**Figure 3.** Field dependence of (a) $\rho_{xy}$ and of (b) $\Delta\rho_{xy}$ for T' BFO at 300 K, (c) field dependence of $\rho_{xy}$ at 5 K and (d) $\Delta\rho_{xy}$ at 5 K and 150 K for R' BFO. In (b) and (d), the error bars are smaller than the symbol size.